# SealMates: Supporting Communication in Video Conferencing using a Collective Behavior-Driven Avatar

MARK ARMSTRONG, Keio University, Graduate School of Media Design, Japan
CHI-LAN YANG, The University of Tokyo, Graduate School of Interdisciplinary Information Studies, Japan
KINGA SKIERS, Keio University, Graduate School of Media Design, Japan
MENGZHEN LIM, Meiji University, Graduate School of Arts and Letters, Japan
TAMIL SELVAN GUNASEKARAN, The University of Auckland, Empathic Computing Lab, New Zealand
ZIYUE WANG, Keio University, Graduate School of Media Design, Japan
TAKUJI NARUMI, The University of Tokyo, Japan
KOUTA MINAMIZAWA, Keio University, Graduate School of Media Design, Japan
YUN SUEN PAI, Keio University, Graduate School of Media Design, Japan

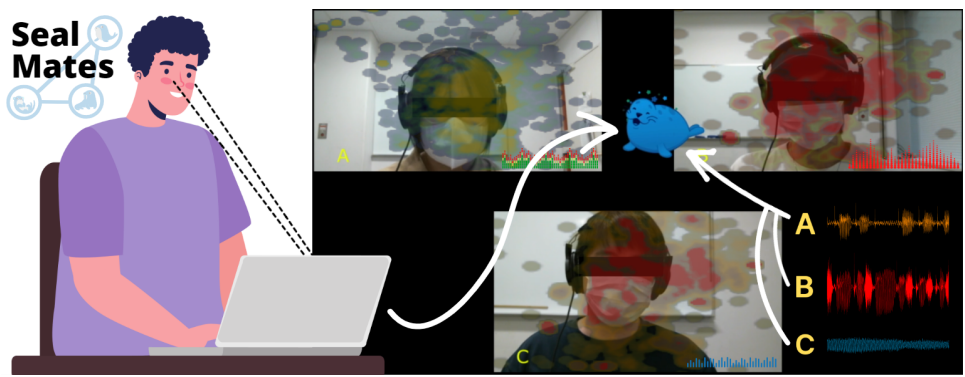

Fig. 1. A heat map of gaze positions (not visible to the participant) and audio activity from all participants are fed to SealMates, which analyzes and changes participants' behavior by redirecting attention towards the participant with the lowest participation level within the triad.

The limited nonverbal cues and spatially distributed nature of remote communication make it challenging for unacquainted members to be expressive during social interactions over video conferencing. Though it enables seeing others' facial expressions, the visual feedback can instead lead to unexpected self-focus, resulting in users missing cues for others to engage in the conversation equally. To support expressive communication and equal participation among unacquainted counterparts, we propose SealMates, a behavior-driven avatar in

118

which the avatar infers the engagement level of the group based on collective gaze and speech patterns and then moves across interlocutors' windows in the video conferencing. By conducting a controlled experiment with 15 groups of triads, we found the avatar's movement encouraged people to experience more self-disclosure and made them perceive everyone was equally engaged in the conversation than when there was no behavior-driven avatar. We discuss how a behavior-driven avatar influences distributed members' perceptions and the implications of avatar-mediated communication for future platforms.

CCS Concepts: • **Human-centered computing** → *Empirical studies in HCI*.

Additional Key Words and Phrases: Avatar-mediated communication, behavior-driven, remote communication, engagement

**ACM Reference Format:**
Mark Armstrong, Chi-Lan Yang, Kinga Skiers, Mengzhen Lim, Tamil Selvan Gunasekaran, Ziyue Wang, Takuji Narumi, Kouta Minamizawa, and Yun Suen Pai. 2024. SealMates: Supporting Communication in Video Conferencing using a Collective Behavior-Driven Avatar. *Proc. ACM Hum.-Comput. Interact.* 8, CSCW1, Article 118 (April 2024), 23 pages. https://doi.org/10.1145/3637395

## 1 INTRODUCTION

Video-mediated communication (VMC) has been widely adopted for distributed people to form various social connections, such as formal discussions and casual chats. However, the insufficient social cues people receive over VMC can make it challenging for remote interlocutors to have equal participation in the multiparty conversation. For example, a lack of mutual eye contact and posture information made it hard for people to find the right timing to manage turn-taking [7, 19]. The absence of cues such as body orientation or physical proximity may add extra difficulties in promoting self-disclosure and building rapport with peers [47].

To support equal participation and encourage self-disclosure in online communication, past works have focused on emphasizing social stimuli and ameliorating exhaustion and awkwardness by changing the visual representation of the communicator. For instance, changing the gaze of the communicators' profile pictures automatically [42, 54] or enhancing the communicators' smile by changing their facial expressions in video conferencing [49]. However, in addition to changing people's visual representations, little has been explored about introducing an external artifact, which is driven by the behavioral data of the group, to mediate and support rapport building in video conferencing. Therefore, we proposed a behavior-driven avatar that can redirect remote interlocutors' attention based on the detected engagement of each member in the group communication.

A behavior-driven avatar has been identified to support expressive communication in text-based communication effectively. For example, Liu *et al.* [23, 24] introduced a behavior-driven avatar that detects physiological change at an individual level to support remote communication among romantic distributed partners. Although a behavior-driven avatar has shown its potential in supporting remote communication, it is still unclear whether and how a behavior-driven avatar that detects the physiological change in a *collective level* shapes the group dynamic differently in video-based communication.

To fill in the research gaps, we aim to address the following three research questions (RQ):

RQ1: How does introducing a behavior-driven avatar impact *self-disclosure* in casual conversation and formal discussion?

RQ2: How does introducing a behavior-driven avatar affects the *perception of equal participation* among members in casual conversation and formal discussion?

RQ3: How does introducing a behavior-driven avatar affect *the actual level of equal participation* for each group in casual conversation and formal discussion?

To answer these three RQs, we proposed SealMates, a behavior-driven avatar within a video conferencing platform to improve remote participants' communication based on their collective engagement. The avatar gathers each participant's speech frequency and gaze patterns by considering its proportion in collective data and then triggers movements and animations across interlocutors' windows. We conducted a controlled experiment to compare the use of (i) behavior-driven avatar, (ii) randomly moving avatar, and (iii) no avatar as a baseline comparison to examine whether communication quality can be changed in terms of self-disclosure and equal participation. We defined and compared the two most common discussion contexts while using VMC: (i) casual conversation and (ii) formal discussion. Casual conversations are typically less structured and more relaxed compared to formal conversations. They may involve exchanging personal information, socializing, or non-work-related topics. On the other hand, formal discussions are structured and organized conversations or meetings that have a specific purpose and follow a set of rules or protocols. Participants in formal discussions engage in a more professional or business-oriented manner and work towards achieving specific objectives within a defined timeframe. Results showed that the behavior-driven avatar encouraged the participants to have more self-disclosure and promote equal engagement in video-based communication. Finally, we discussed design implications regarding the future of avatar-mediated communication for supporting multi-party video-based communication.

Based on the empirical data, the two main contributions of this work are:

(1) We propose and evaluated SealMates, a behavior-driven avatar system in video conferencing, which can enhance levels of self-disclosure and perceived equal participation during casual conversation over video conferencing
(2) We expanded the definition of avatar-mediated communication by using *collective* behavioral data to facilitate group interaction through avatar movement.

## 2 RELATED WORK
### 2.1 Current Challenges in Video Conferencing

Video conferencing enables remote individuals to communicate while observing facial expressions and transmitting audio. However, compared to in-person interaction, video conferencing still lacks several crucial social cues that play a vital role in establishing connections and building rapport. Drawing upon extensive research in video-mediated communication, we have identified two significant issues with the current video conference systems that require further attention to promote more expressive and engaged video conference experiences. These issues include (i) potential distractions caused by seeing one's own visual feedback [12], and (ii) difficulties in managing turn-taking [34].

Firstly, the camera function allows meeting users to watch not only other users' video feed but also their own. A study using an eye tracker showed that in interactive meetings between small groups, participants looked away from the screen about 33% of the time and looked at their own video frequently [12]. At the beginning of a meeting, people may want to check their appearance by focusing on their own video [6]. Goffman [14] found that conventional conversation with the lack of self-perception generally plays out smoothly. A video conference platform that allows user to see themselves increases self-awareness and self-consciousness [27], and this could impact the comfort level and hinders effective communication [6]. Users' attention is also easily influenced when they are allowed to focus on themselves, which can often lead to negative feedback loops of self-focus. The ability to watch their own video also affects their comfort level and hinders effective communication [6]. In the job interview context, it increased one's cognitive load [27] and can also lead to Zoom Fatigue [30].

Secondly, during a face-to-face interaction, people often indicate their intention to take the floor or leave it to others with non-verbal cues such as eye contact. However, the absence of mutual gaze in video conferencing poses a challenge in effectively managing turn-taking, often resulting in unequal participation among participants. Several approaches have been suggested to address this issue in video conferencing, such as the use of generated gaze [54], turn-taking prediction algorithms [18], and the addition of extra communication channels and visualization to facilitate balanced participation [41]. These existing approaches focus on providing explicit design interventions to change people's behavior, including showing participation percentages or gaze direction. However, there is a lack of understanding when it comes to subtle interventions that can change human perception. Therefore, we propose a behavior-driven avatar that is expected to dynamically shift everyone's attention through its movement based on individual engagement levels within the group. It will identify and respond to each participant's degree of involvement, with the goal of enhancing turn-taking dynamics and promoting equal participation in video conferencing. By considering the group as a whole and adapting the avatar's behavior accordingly, we aim to enhance the overall turn-taking dynamics and promote equal participation in video conferencing.

## 2.2  Augmenting Social Cues Based on User Behavior in Video Conferencing

Using biosignals to augment social cues is one example of how user behavior has been leveraged to support social interaction in video conferencing. For instance, Wang *et al.* [52] proposed a chat system where the text was animated to reflect the senders' emotional state obtained via galvanic skin response. With this design, receivers could better sense the socio-emotional cues from the senders. Furthermore, Tewell *et al.* [50] found that the temperature change can be used to convey emotion in text messages and can change the level of perceived arousal of text messages. Additionally, the visualizations effectively conveyed the various levels of avatar arousal as indicated by the biosignals [22]. Experiments on deception games observed that the presence of biosignals, such as visual displays of heart rate (HR), had a significant impact on players' strategies and influenced their gameplay [9]. In remote video game settings, when both bio- and video information were made available simultaneously, the sense of social presence closely mirrored the baseline condition of players sharing the same physical space [16]. However, the above approaches require users to wear an array of sensing devices, which is challenging for scaling up the interaction across various scenarios.

Gaze and utterances are two major channels for people to manage turn-taking in communication and can be easily captured in video conferencing [41]. This is because participants in video conferences primarily direct their attention towards others[37]. Simultaneously, verbal cues data can be used to predict who will be the next speaker [28]. Therefore, we aim to mediate remote communication based on the gaze and speech data from the distributed members so that they are not required to use any wearable devices during video conferencing.

## 2.3  Avatar-Mediated Communication (AMC)

Avatar-mediated communication (AMC) was broadly defined as using a digital representation of a human user to interact with other users in an online social setting [31]. In this study, we aim to expand the definition of AMC from one digital representation of a user to one digital representation of a group of users in computer-mediated communication. A rich literature has shown that introducing an avatar as a user representation and an avatar as an external mediator in computer-mediated communication influence group dynamics and communication quality [8, 24, 33]. In terms of using avatars as a user representation, related works have shown that changing the facial expression of the senders' avatar could lead receivers to have a positive impression of the sender [33], or using avatars to reduce social anxiety in an online debate [35]. In terms of using an avatar as an external mediator, a prior study has demonstrated a biosignal-driven avatar could support authentic communication between romantic partners [24]. Although it has been demonstrated that an avatar driven by a single user can influence online communication, it is unclear how an avatar that is driven by *collective behavioral responses* could influence group dynamics. In video conferencing, where people lack nonverbal cues such as mutual eye contact and proximity information, it can be difficult for people to manage turn-taking, harming rapport building and equal participation. Hence, this study aims to explore how a collective behavior-driven avatar can serve as a visual guide to facilitate turn-taking in video conferencing.

## 3 SYSTEM DESIGN

Our system objective is to distribute participation among video conference participants equally. To achieve this, the first step is to identify a metric by which user participation can be measured in a conversational scenario; a Participation Score. With this metric, we can introduce a new factor into the environment and perform an analysis to check for any significant difference in recorded participation. The following section will describe in detail our participation metric system and the considerations that led to this model. The design considerations and decisions for this system stem from the synthesis of previous literature continuous informal pilot testing as well as feedback we gathered.

### 3.1 Design Rationale

*3.1.1 Scoring.* The score is divided into the areas of participation that we aim to address with this system. Since sensory perception in most video conferences is strictly limited to audio and video, these are the two channels for data collection, visualized in Figure 1. Previous literature identifies problems with self-focus and lack of attention to other meeting participants [3, 14], so the visual score will reflect the average amount of time all participants in the meeting spent looking at a single user, using eye-tracking for data collection. While speaking too much or too little can also lead to ZOOM fatigue [5, 17], the audio score will reflect how much time a participant has spent speaking, using microphone activity for data collection. In both cases, an ideal score that reflects equally distributed participation from a user would be 1/N where N is the number of meeting participants. As N becomes large, the meeting will become too difficult to distribute even participation among all members, so this system design is for casual and small video conferences with three participants in this system design.

*3.1.2 Avatar Design.* We propose to use an animated avatar to correct attention and distribute engagement, which can be applied across disciplines to designers for Computer-Mediated Communications. Previous work demonstrated that an avatar with multiple animation states in a telecommunication context can be used to improve social connection between users [23]. Additionally, two key features that drive our avatar design are the inclusion of multiple animation states and whether or not it semantically fits the communication context [24]. Therefore, following the design style of previous works, our system will use an aquatic-style avatar that can navigate in all directions in a 2D environment, similar to water. The avatar has several ambiguous animation states that can be interpreted by users in different ways and will be described in the SealMates Features section.

## 3.2 Participation Score Formula

It can be argued that sound or video are equally important depending on user preference. For a limited-use, evenly-weighted score, we consider audio and video elements to consist of equal importance (50% each) in our final participation score calculation.

First, our system analyzes the microphone activity of each user at 60Hz, with a volume threshold to exclude insignificant background noise. Each frame where the mic volume is above the threshold is summed over one minute (3600 frames), resulting in a percentage, the Audio Participation Score (APS), which is half of the total Participation Score of A.

Secondly, the Visual Participation Score (VPS) reflects how much a user participated in the conversation as a measurement of how much their window was looked at. This section involves the use of gaze tracking and is accounted for in two different ways: self-scores and collective scores. To measure how long user A looked at themselves, we calculate the sum of frames where A's gaze overlapped A's bounding box over one minute (3600 frames). This percentage is the self VPS of A from A. How long user A looked at user B, would result in the self VPS of B from A.

A single user A will produce 3 self VPS, as will users B and C. Therefore, to calculate the collective VPS of user A, we average all self VPS on A (from A, B, and C). The result is half of the total Participation Score for user A.

## 3.3 Hardware

The data collection is performed by a webcam microphone[1], and a commercial use eye tracker (in our case the Tobii EyeX and Tobii 4C eye trackers[2]). It should be noted that these external devices are not included in most video conference platforms. However, it can promote the development of better webcam gaze-tracking algorithms for future work. Our software (TouchDesigner[3]) records data at a rate of 60 frames per second, so audio was evaluated based on the number of frames that passed the volume threshold in the time window.

The eye trackers transmit an eye gaze coordinate via User Datagram Protocol (UDP) that is remapped to the size of our device screen, a 15" laptop. The eye gaze coordinate is recorded at our sample rate (60Hz) and evaluated at each frame to see if the point lies within the bounding box of each user. The bounding boxes that surround each user will be classified as Areas of Interest for gaze data analysis. If the gaze point of Participant A is inside of the bounding box of Participant C for example, then it is recorded for that sample that "Participant A was looking at Participant C". If the gaze point is not within any bounding box, off-screen, or if the data is undetected, the sample is classified as "Participant A was looking at Nothing". Additionally, we also check to see if there is an overlap between the gaze point and the bounding box of our avatar. For every sample where there is overlap, it is also recorded that "Participant A's gaze was overlapping with the avatar". This is classified separately from window attention because the bounding box of the avatar is non-stationary, which we believe might provide a correlation between avatar location and user window attention scores.

## 3.4 SealMates Features

*3.4.1 Initial Testing.* The stylistic design of our system is influenced by the work 'Significant Otter' [24] in which an aquatic avatar is used for remote physiological signal-based interaction between users. Our initial avatar representations included a procedural 3D Jellyfish, which we found to lack the emotional expression necessary to capture attention, and a 2D Rabbit, which did not have enough different animation states to represent intelligence. We found and settled on a series of six free-use seal GIFs[4] (Figure 2) which could represent one of three different animation states, using two unique animations so that users wouldn't recognize the repetition. The animation states are idle, moving, and a precursor to movement animation that implies contemplation of the avatar.

---

[1] https://www.logicool.co.jp/ja-jp/products/webcams.html
[2] https://gaming.tobii.com/
[3] https://derivative.ca/

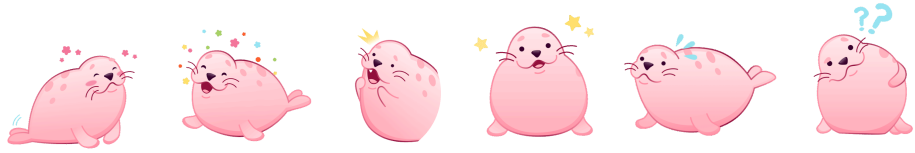

Fig. 2. Six unique seal animation state GIFs, from left to right: Idle-1, Moving-1, Precursor-1, Idle-2, Moving-2, Precursor-2. The GIFs were adopted from: https://tenor.com/search/seal-stickers-stickers

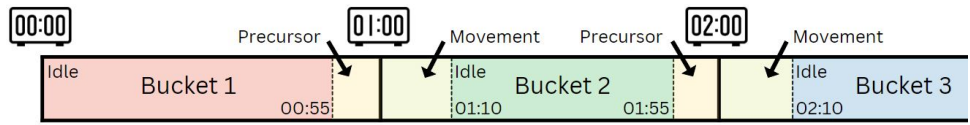

Fig. 3. Each time bucket is 1 minute in length, and at each mark, the avatar calculates where to move next. 5 seconds before each movement it switches to a precursor animation to indicate that it is thinking, followed by a 10 second movement animation as it moves to its next position. The remaining time, the avatar is idle.

*3.4.2 Avatar Behavior.* During a video conference, the avatar should move at a pace that is not too frequent to cause distractions but also in a meaningful way that can feel like a natural event, drawing the user's attention. We divide a video conference into temporal buckets so that the avatar can analyze activity continuously without its decision-making process being weighted by sliding window decay. After pilot testing with 14 dyads of participants, we decided on 60s time division because frequent avatar movement was less noticeable to pilot testers than 30s divisions and more active than 2m divisions.

The avatar starts off in the middle space between all three participant video bounding boxes and will spend one minute recording data from all three users in the form of gaze points and audio activity. Five seconds before the end of a minute bucket (see Figure 3), the avatar switches to a precursor animation state, which implies that it is thinking about where to move next. When the avatar follows our algorithmic-behavior design for choosing where to move next (outlined in Figure 5), it will be referred to as an Algorithmic-Avatar, or alternatively, a Behavior-driven Avatar. For this behavior, at the end of each one-minute segment, all Audio Participation Scores are calculated using Equation 1 (detailed later in the Measurements subsection), as well as the collective VPS for each user through Equation 2. The total Participation Score becomes a percentage out of 100, following our behavior-driven formula. After discovering the user with the lowest score, the avatar switches from its precursor animation to a moving animation, where it will move across the screen over a period of ten seconds from its current position to the currently lowest score user. In the case of a Random-behavior avatar, it simply chooses a random location in a bounding box to move to. After this movement period, the avatar finally switches back to its least recent idle animation, and the cycle continues for the next one-minute bucket. If the avatar does not intend to move, it skips the movement animation. These ambiguous behaviors are open to interpretation as the goal is not to outcast a low-score participant but to redirect attention through onscreen movement.

---
[4]https://www.baretreemedia.com/

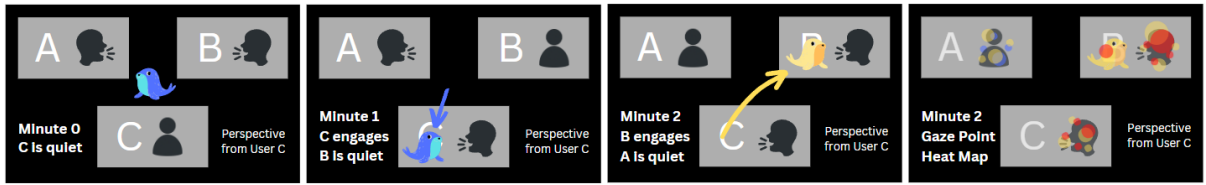

Fig. 4. The avatar spends the first minute collecting data from users in the neutral middle of the screen (left). In a hypothetical scenario, A and B are engaging a lot with each other and ignoring C, so after one minute the avatar moves to C. During this next minute (middle-left) A starts to talk with C, where B becomes less engaged. In the following minute (middle-right), the avatar migrates to B, and its color from C's perspective shifts yellow, a result of C's increased audio score. While the avatar is with B, C's fixations may be on themselves and B, with fewer on A (right).

The avatar also features some color variation which is unique to each individual screen (see Figure 4). Depending on the Audio Participation Score from each participant, the avatar will shift colors from blue (at low scores) to yellow (at high scores). This color is different to all users in the case that if the avatar becomes the subject of discussion, the variation in visual presentation may defer participants from inferring its functionality.

### 3.4.3 Expected Interaction Examples.
We discuss potential conversational scenarios using Seal-Mates. In these scenarios, we envision a triad setting similar to Figure 4, where the participation and the attention among the participants are less than ideal:

(1) In a conference situation where participant A becomes too chatty, and B and C have low Audio Scores, participant A might see a yellow seal and make a comment on it, while B and C agree that the seal is in fact blue on their end. This point of contention might cause B and C to speak more.
(2) In another situation A and B spend much time focusing on B, while C stares at C. A's participation score might be drastically lower which would prompt the avatar to move to A. We expect the avatar would direct more attention to A, changing the focus behavior of all users in the conference.
(3) In one situation, User C is focusing too much on themselves and not speaking a lot. Their self-focus would be high but won't pollute the data because it is averaged with other participants. Meanwhile, their audio score is low. The avatar may move to another user, which we expect would redirect the gaze of C. Or the avatar may move to C, prompting C to think they are being watched and that they should participate more.

## 4 USER STUDY

The goal of the study is to investigate whether video conferencing with a behavior-driven avatar will have different perceptions and behavioral tendencies than using a random avatar or without the presence of an avatar.

According to previous research showing animated avatars enabled expressive communication [23], we hypothesized H1a and H1b:

**H1a:** Having video conferencing with SealMates will lead people to have more *self-disclosure* than using a random avatar or no avatar *in casual conversation.*

**H1b:** Having video conferencing with SealMates will lead people to have more *self-disclosure* than using a random avatar or no avatar *in formal discussion.*

Next, it has been demonstrated that visualizing conversational behavior, such as frequency of interruption and turn-taking, improved equal participation in video conferencing [10]. Furthermore,

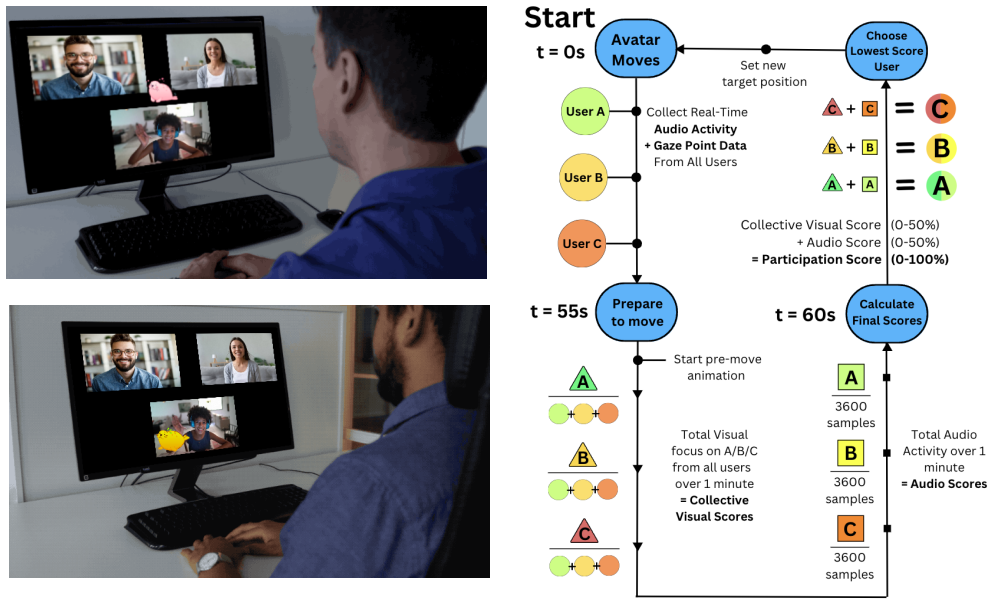

Fig. 5. Beginning in the middle of the screen (top left), the avatar's algorithmic decision-making process is demonstrated in a continuous loop (right), where it calculates at the near-end of each time bucket, each user's participation score, and sets its destination position to the bounding box of the least active user, and moves there (bottom left).

people's perception of participation might be different from the actual participation in a group conversation. Hence, we hypothesize H2a and H2b based on participants' perception, while formed H3a and H3b based on their participation level inferred from their behavior:

**H2a:** When having video conferencing for *casual conversation*, showing SealMates will make people *perceive everyone participating equally* in the conversation more than using a random avatar or no avatar.

**H2b:** When having video conferencing for *formal discussion*, showing SealMates will make people *perceive everyone participating equally* in the conversation more than using a random avatar or no avatar.

**H3a:** When having video conferencing for *casual conversation*, showing SealMates will lead to *equal participation* in the conversation more than using a random avatar or no avatar inferred from speech amount and visual engagement.

**H3b:** When having video conferencing for *formal discussion*, showing SealMates will lead to *equal participation* in the conversation more than using a random avatar or no avatar inferred from speech amount and visual engagement.

### 4.1 Participants

To gather relevant insight from a demographic of users who actively use video-conferencing platforms on a day-to-day basis, we recruited 45 participants (30F, 14M, and one prefer not to say) of international and local ethnicity, with an average age of 26.38 years old (SD = 5.14) by word-of-mouth campaign and social media posting at an Asian university. Participants were invited to meeting rooms within the authors' campus to join a small group discussion task involving video conferencing with two other unacquainted partners. Conversational English proficiency, adulthood, and regular prior video-conference experience were the only prerequisites for entry. They were paid approximately $20 USD for 75 minutes of participation. The amount was determined based on

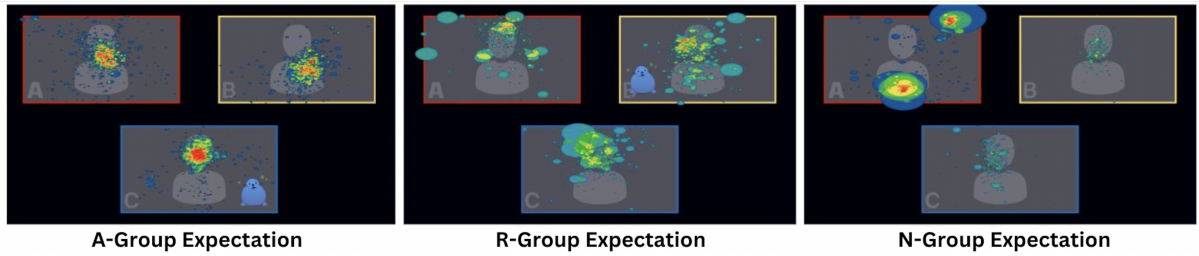

Fig. 6. We expect that the Behavior-driven avatar condition (left) will demonstrate features of more equalized fixation distribution across all users, with high fixation counts centered on the subjects who are expected to engage more when sharing a window with the avatar. Random Avatar condition behavior(middle) is expected to create a similar number of fixations but a non-equal distribution across participants as some may share a window with the avatar for too long, leading to more sporadic eye movement from boredom. No avatar condition (right) fixation behavior is characterized by a generally lower fixation count, non-homogeneity across all users, as well as the possibility of long-duration fixations on distractions in the scene and lower-duration fixations on some participants.

the standard hourly wage in the country where the study was conducted. The ethical review board of the authors' institute approved the study.

## 4.2 Experiment Design

We assigned participants to three conditions based partially on availability while also taking into account participant familiarity. Every condition consists of three participants. In total, we formed five triads in each condition. In our ideal scenario, participants would treat each other as professional colleagues. The intended application of this work would be to use our system to improve the participation of users in environments where participation is usually mandatory but not enthusiastic. Almost all triads (save one) included at least one or more members who were unfamiliar with the other members on a close-friend basis. To ensure that all triads did not socially outcast any participant, every triad was instructed beforehand to act professionally towards other participants, as if in a workplace setting, and to avoid any inside jokes or preferential treatment towards any specific partner.

Our experiment followed a between-subjects design. We simulated two common workplace scenarios, including casual conversation and formal discussions. Our independent variable in this case was types of avatars, which includes *Behavior-driven avatar, Random avatar*, and *No avatar* conditions. The first, "Behavior-driven" was the condition where the avatar would behave based on our proposed algorithm, creating a participation score and driving the avatar to move to the areas of the screen with the lowest-score participant. The second condition, "Random avatar," was the only condition other than A, that involved an avatar. The avatar in this condition followed no specific pattern of movement and would choose at random what location on the screen to move to at each interval. The final condition, "No Avatar" was the control condition, where any avatar-related data or information was omitted from materials.

To answer each research question and hypothesis, we had the following dependent variables: The first of which is the perceived level of self-disclosure followed by the willingness to have further interaction. **(RQ1, H1)** Next is the perception of equal participation from each user **(RQ2, H2)**. Lastly is the raw log data of gaze points and audio activity of each participant, which will measure participation according to our proposed participation score formula calculated in Figure 5, as well as the number of fixations, saccades, and patterns in behavior derived from the log data will be analyzed across each condition **(RQ3, H3)**. Examples of our expected fixation behavior patterns are listed in Figure 6.

### 4.3 Procedure

Upon arriving at the study, participants were given a brief explanation of the nature of the experiment, which was to collect physiological data from users while participating in casual conversations over a teleconference platform, and were instructed to move into isolated rooms, visualized in Figure 7. Once in their respective rooms, they were shown a three-minute video explaining their withdrawal rights, what methods of data collection we would use, how long the study should take, and how many activities they would participate in. If they agreed to participate in the study, they were presented with a participation consent form.

Next, the participants were prepared for eye tracker calibration. We used the Tobii Eye Tracker models 4C and EyeX. The eye tracking calibration process takes about 5 minutes, as we would calibrate once in the Tobii Software and once more in our custom program to ensure the gaze point accuracy.

After calibration, the participants were instructed about their first casual chat task. The experimenters asked them to keep all topics discussed in this conversation strictly confidential, even after the study. They were given six minutes to hold a casual chat about the two following prompts based on [2]: *"Where in the world do you want to travel?" and "What superpower would you like to have?"*. Conditions with the avatar were informed that there would be an avatar on the screen who did not speak but would join in on the discussion as an observer.

After the six-minute casual chat, participants were presented with a survey asking their perception of the conversation. Once they were finished with the survey, they moved on to the formal discussion task, which was a role-play negotiation.

In this role-play negotiation, participants were instructed to persuade their team members based on their assigned roles. Participants were instructed that they had to plan an end-of-year party for their shared company. Each participant was given a role as Head of Security, Finance, or Social Events. With each role came four unique primary objectives, which they were tasked to convince their colleagues to agree on so that they could earn a salary bonus in the company. The four objectives to agree on were the ticket price to bring friends to the party, how many entertainers to hire, how many security guards to hire, and what time the party should end. Participants were informed that they would have 15 minutes for this activity. The task was adopted by [1].

After being presented with their role and negotiation objectives, participants were given an Understanding Check survey to record if they understood their roles. The experimenter would check again for eye calibration accuracy and visual or audio delay before this activity began. After this formal discussion task, participants were given the same post-survey and asked about their perception of the conversation. Upon completing the survey, participants were given their monetary reward and allowed to ask any question about the study.

### 4.4 Measurements

*4.4.1 Self-disclosure.* To evaluate how SealMates affects the degree of self-disclosure in the remote conversation, we adopted the survey from Sprecher [46]. We asked participants to rate the following four questions after each round of conversation: *How much did the group tell you about himself or herself? How much personal or intimate information did the group share with you? How honest and open do you think the group was in their responses? How much knowledge do you think you gained about the group?* The survey was conducted with a seven-point Likert scale, where one indicated "Not at all" and seven indicated "A great deal". The average score from the four items served as an index for self-disclosure.

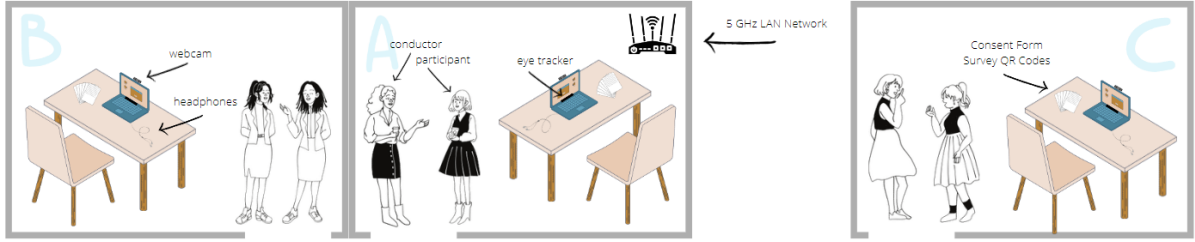

Fig. 7. Study layout. The study was conducted in 3 rooms.

*4.4.2 Perceived Level of Equal Participation.* After each round of conversation, we also instructed participants to indicate who they thought was the *most* and the *least* engaged person in the conversation. If participants reported one of the other two members as the most or least engaged member, then we assigned "other members" as a label for data analysis. If participants reported three of them had similar levels of engagement, then we labeled it as "equal". When participants indicated themselves, then we labeled it as "self" for Chi-square analysis. (Figure 9 conversation and Figure 10) We designed this question to examine whether SealMates influenced the perceived participation level of the remote counterparts.

*4.4.3 Engagement Distribution Inferred by Audio and Gaze Data.* During the interaction, we recorded the log data, which can derive the duration and amount of speech as well as the duration of gaze in each window for each participant. To examine whether SealMates changes the level of participation in the discussion, we quantify the oral and visual engagement of each user with the following two indexes by referring to [43]:

$$LevelOfCollectiveOralParticipation = \frac{\sum_{t=15}^{t=1} APS_{Xt}}{\sum_{t=15}^{t=1} APS_{Wt}} \quad (1)$$

Where t is each minute, X is the currently observed person, W is the union of all users' audio activity, and APS is Audio Participation Score.

$$LevelOfCollectiveVisualParticipation = \frac{\sum_{t=15}^{t=1} XVPS_t}{\sum_{t=15}^{t=1} WVPS_t} \quad (2)$$

Where t is each minute, VPS is the Visual Participation Score, X is the currently observed participant, and W is the VPS of all participants combined.

For both indexes, we will get a percentage score ranging from 0 to 1. A higher score means more equal contribution. We calculated the audio and visual participation scores from each group after a casual conversation and formal discussion across all conditions.

To examine the visual attention and fixation on AoI, we utilize log data to extract measures including fixation start time, fixation end time, fixation duration and corresponding gaze x, y coordinates in the screen, total fixation count, total fixation duration, average fixation duration, and fixation count per minute related to attention on the screen's AoI for all the participants. We followed the fixation analysis pipeline from raw data and implemented it in Python using PyTrack-NTU Library [13] and extracted the above fixation features. We set fixation thresholds at 150 milliseconds of stable fixation and extracted the list containing the time fixation began, its end, duration, and coordinates.

We calculated the fixation count and duration on each AoI for each participant when the avatar was present in the window when the avatar was not the AoI and for the whole duration of the

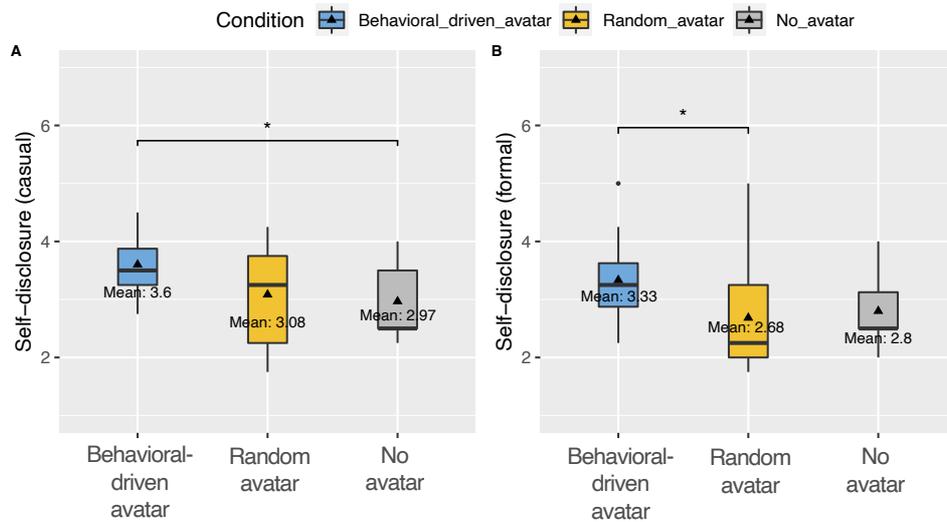

Fig. 8. Level of perceived self-disclosure after performing a casual (A) and formal (B) discussion shown in a box plot. The horizontal black line in the box denotes the median, whereas the black triangle denotes the mean score. The y-axis shows participants' perceived self-disclosure level from other members. A higher score indicates more self-disclosure participants perceived. The x-axis shows all conditions compared in the study. The blue box represents the condition of a behavior-driven avatar (SealMates), the yellow box represents the condition of a random avatar, and the gray box represents the condition of no avatar. Results demonstrated that SealMates made participants perceive higher self-disclosure from their remote partners than when there was no avatar in a casual conversation (A) and when using a random avatar in a formal discussion (B). * denotes p-value < 0.05

meeting. Our primary measures of interest, classified as Behavioral Feature Categories, are the following:

(1) Relative Fixation Duration (RFD) was calculated by determining the proportion of time spent fixation at the AoI out of the total meeting time.
(2) Relative Avatar-Fixation Duration (RAFD) was calculated by determining the proportion of time fixation at the AoI when the avatar was present at the AoI, out of the total fixation duration at this particular AoI.
(3) Relative Fixation Count (RFC) was calculated to determine the proportion of fixation counts per AoI, out of the total fixation counts of all AoIs during the meeting.
(4) Relative Avatar-Fixation Count (RAFC) was calculated to determine the proportion of fixation counts when looking at the AoI while the avatar was present in AoI, out of the total fixation counts.
(5) Relative Audio Score (RAS) was calculated by determining the proportion of audio activity by the participant out of the total proportion of meeting time when there was ongoing audio activity. See Equation 1.
(6) Relative Minute to Minute Fixation Score (RM2MFS) was calculated to determine the minute-to-minute proportion of Maximum fixation counts by a participant when looking at the AoI while the avatar was present in AoI, out of 15 minutes.

## 5 RESULTS

### 5.1 Self-disclosure (RQ1, H1)

A Shapiro-Wilk test did not show evidence that the data fits the normality assumption, thus, non-parametric statistics was used for analysis. A Kruskal-Wallis test on casual conversation showed that the presence of an avatar in video conferencing had a significant effect on self-disclosure (**H1a was supported**, $H[2] = 5.35$, $p = 0.068$, $\eta_p^2 = 0.08$, moderate effect). Post-hoc testing using a Benjamini-Hochberg method for adjusting p-value showed that the people disclosed more about themselves when using a behavior-driven avatar (Mdn = 3.5) than without any avatar (Mdn = 2.5) ($p < .05$) (Figure 8A). Similarly, a Kruskal-Wallis test on formal discussion showed that the presence of an avatar in video conferencing had a significant effect on self-disclosure ($H[2] = 7.90$, $p < 0.05$, $\eta_p^2 = 0.14$, large effect). Post-hoc testing showed that people disclosed more about themselves when using a behavior-driven avatar (Mdn = 3.25) than using a random avatar (Mdn = 2.25), and no avatar (Mdn = 2.5) (Figure 8B).However, behavior-driven avatar did not make people disclose significantly more to their remote members compared to no avatar condition (**H1b was not supported**)

### 5.2 Perceived Level of Equal Participation (RQ2, H2)

We compared how participants in every condition perceived the most and the least engaged members in the group. When asking participants to indicate who was the most and the least engaged person in the conversation, the Chi-square result showed that for both casual and formal discussion, the behavior-driven avatar made people perceive more equal engagement of each member compared with the situation when there was no avatar.

After having a casual conversation, participants in no avatar condition, significantly more participants reported that other members were engaged and not engaged in the discussion than equal participation (Reporting the most engaged person: $X^2(2) = 11.2$, $p < .01$; Reporting the least engaged person: $X^2(2) = 11.2$, $p < .01$). However, such polarized perception was not found in behavior-driven avatar condition when reporting the least engaged members (Reporting the most engaged person: $X^2(2) = 7.6$, $p < .01$; Reporting the least engaged person: $X^2(2) = 1.6$, n.s.). (Figure 9, **H2a was partially supported**)

A similar result was found for formal discussion. More participants in the no avatar condition reported other members were engaged or not engaged in the discussion than equal participation (Reporting the most engaged person: $X^2(2) = 14.8$, $p < .01$; Reporting the least engaged person: $X^2(2) = 12.4$, $p < .01$). However, participants in behavior-driven avatar did not have such polarized perceived engagement after discussion (Reporting the most engaged person: $X^2(2) = 5.2$, n.s.; Reporting the least engaged person: $X^2(2) = 1.2$, n.s.). (Figure 10, **H2b was supported**)

### 5.3 Engagement Distribution Inferred by Audio and Gaze Data (RQ3, H3)

To gain further insights into participants' gaze behavior and their areas of interest (AoI), We then performed analysis on the gaze behavior of every participant, with a focus on fixations, to understand the nature of participants' Areas of Interest (AoI), and how often they were centered around the avatar and compared them among different avatar conditions. A Shapiro-Wilk test on RAS (casual and formal discussions), RAFC, RAFD, RFC, RM2MFS and RFD (refer to subsubsection 4.4.3 for the detail of each parameter) across all conditions did show evidence that the data fits the normality assumption. Thus, parametric statistics were used for analysis. As per our experiment design, our independent variables are the type of avatars (with three levels: Behavior-driven, Random avatar, and No Avatar), and the dependent variables are participants' behavior due to the avatar's presence on the gaze attention data and audio score. Hence, we carried out one-way ANOVA and Tukey HSD for post hoc analysis of these variables.

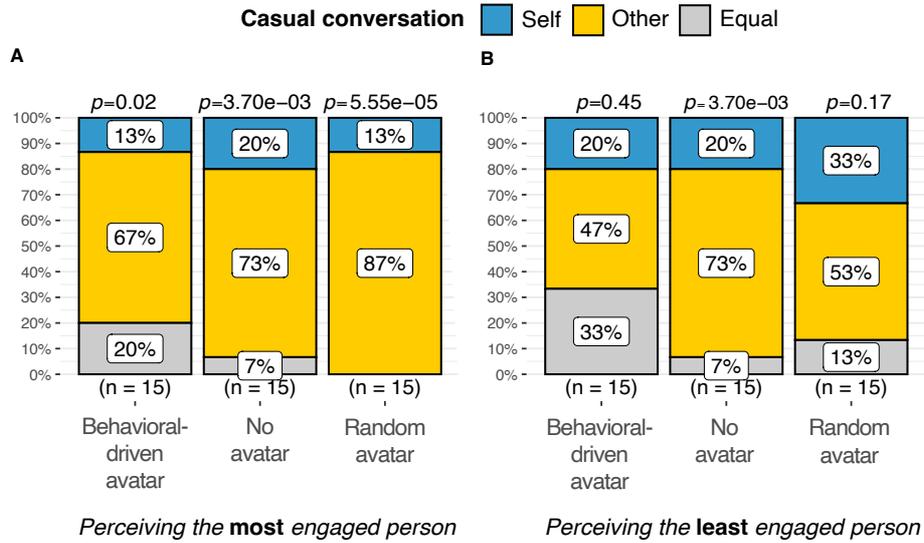

Fig. 9. Perceived equal participation after casual conversation

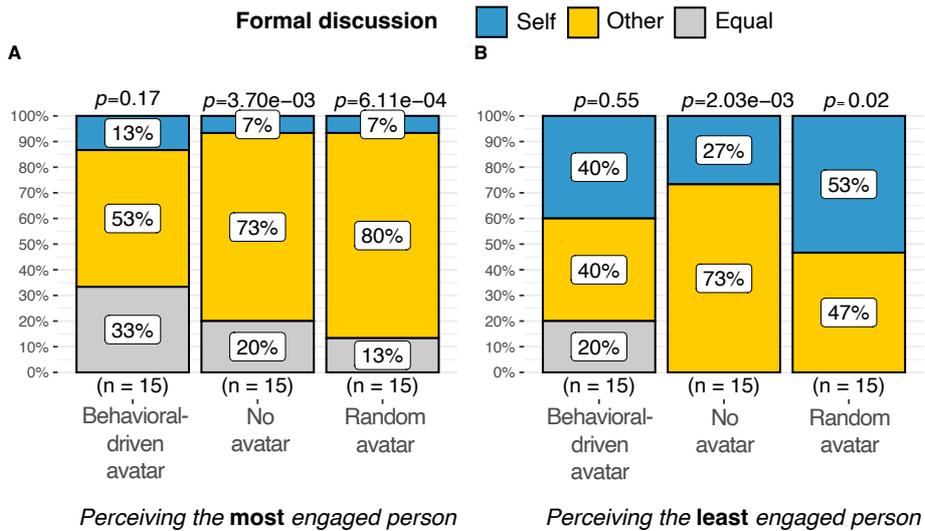

Fig. 10. Perceived equal participation after formal discussion

*Relative Audio Score (RAS).* One-way ANOVA showed that there was a significant difference between behavior-driven avatar condition, random avatar condition, and no avatar condition on RAS in casual conversation ($F[2,42] = 7.127$, $p = 0.002$). Further post hoc pairwise comparisons using Tukey's HSD showed that the behavior-driven avatar condition ($M = 57.42, SD = 9.3$) led to a higher RAS mean than the random avatar condition ($M = 21.2, SD = 12.4$) and no avatar condition ($M = 41.6, SD = 10.5$), respectively (Figure 11a).

Similarly, one-way ANOVA showed that there was a significant difference between behavior-driven avatar condition, random avatar condition, and no avatar condition on RAS in formal discussion ($F[2,42] = 10.846$, $p = 0.000023$). Further post hoc pairwise comparisons using Tukey's HSD revealed that the behavior-driven avatar condition ($M = 48.69, SD = 14.44$) led to a higher

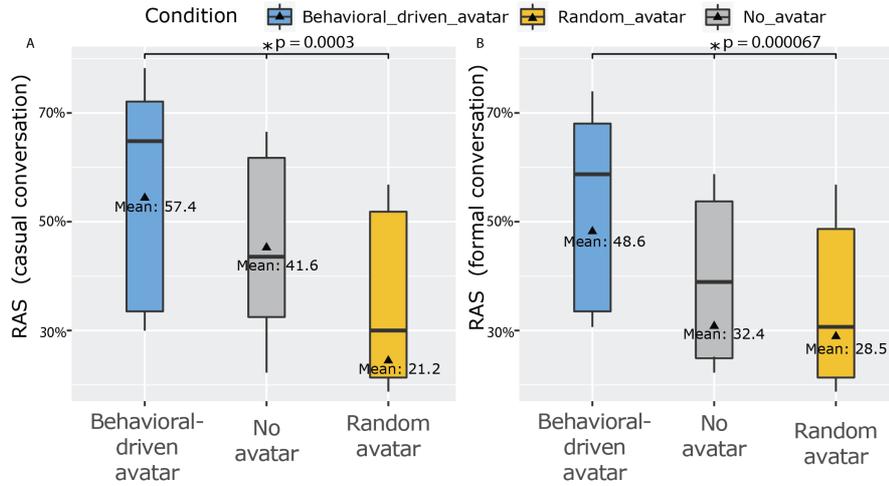

Fig. 11. RAS on casual and formal conversations

RAS mean than the random avatar condition ($M = 28.58, SD = 7.104$) and no avatar condition ($M = 32.71, SD = 9.25$), respectively (Figure 11b).

*Relative Fixation Count (RFC), Relative Fixation Duration(RFD).* One-way ANOVA showed no significant difference across behavior-driven avatar condition ($M = 93.45, SD = 3.164$), random avatar condition ($M = 90.2, SD = 6.632$), and no avatar condition ($M = 92.86, SD = 4.149$) when comparing RFC ($F[2,12] = 1.272$, *n.s.*, Figure 12c). Similarly, there was also no significant difference across behavior-driven avatar condition ($M = 54.36, SD = 16.2$), random avatar condition ($M = 50, SD = 14.26$), and no avatar condition ($M = 47.86, SD = 11.38$) when comparing RFD ($F[2,12] = 0.642$, *n.s.*, Figure 12d).

*Relative Avatar-Fixation Duration (RAFD), Relative Minute to Minute Fixation Score (RM2MFS), Relative Avatar-Fixation Count (RAFC).* Next, we analyzed the difference between the random avatar and behavior-driven avatar condition on the gaze attention data. We performed Student's T-tests for RAFD, RM2MFS and RAFC. Because avatars only appeared in two conditions, Student's T-test was used for analyzing these gaze data and only the data from two conditions were shown in Figure 12a, b, e. However, there was also no statistical significance between behavior-driven avatar condition ($M = 35.97, SD = 9.02$) and random avatar condition ($M = 35.15, SD = 7.49$) on comparison with RAFD ($t(14) = 0.096$, *n.s.*). Likewise, there was no statistical significance between behavior-driven avatar condition ($M = 3.53, SD = 2.01$) and random avatar condition ($M = 3.69, SD = 2.18$) on comparison with RM2MFS ($t(14) = 0.854$, *n.s.*). A similar situation is found between behavior-driven avatar condition ($M = 35.57, SD = 8.32$) and random avatar condition ($M = 35.23, SD = 8.074$) on comparison with RAFC($t(14) = 0.066$, *n.s.*).

Furthermore, we performed a Pearson rank-order correlation test for parametric data between RAS and RAFC, RAS and RAFD, RAS and RFD, and RAS and RFC across all conditions. The result showed a strong positive correlation between RAS and RAFC ($\rho = 0.8756$, $p < 0.001$), RAS and RAFD ($\rho = 0.827$, $p < 0.001$), RAS and RFD ($\rho = 0.7865$, $p < 0.001$), RAS and RFC ($\rho = 0.7858$, $p < 0.0001$) in behavior-driven avatar condition.

In contrast, we found a weak correlation between RAS and RAFC ($\rho = 0.137$, $p = 0.18$), RAS and RAFD ($\rho = 0.036$, $p = 0.52$), RAS and RFD ($\rho = -0.103$, $p = 0.32$), RAS and RFC ($\rho = 0.366$, $p = 0.174$) in random avatar condition. Similarly, we found a weak correlation between RAS and RFD ($\rho$ = -0.243, $p$ = 0.746), RAS and RFC ($\rho$ = 0.179, $p$ = 0.89) in no avatar condition.

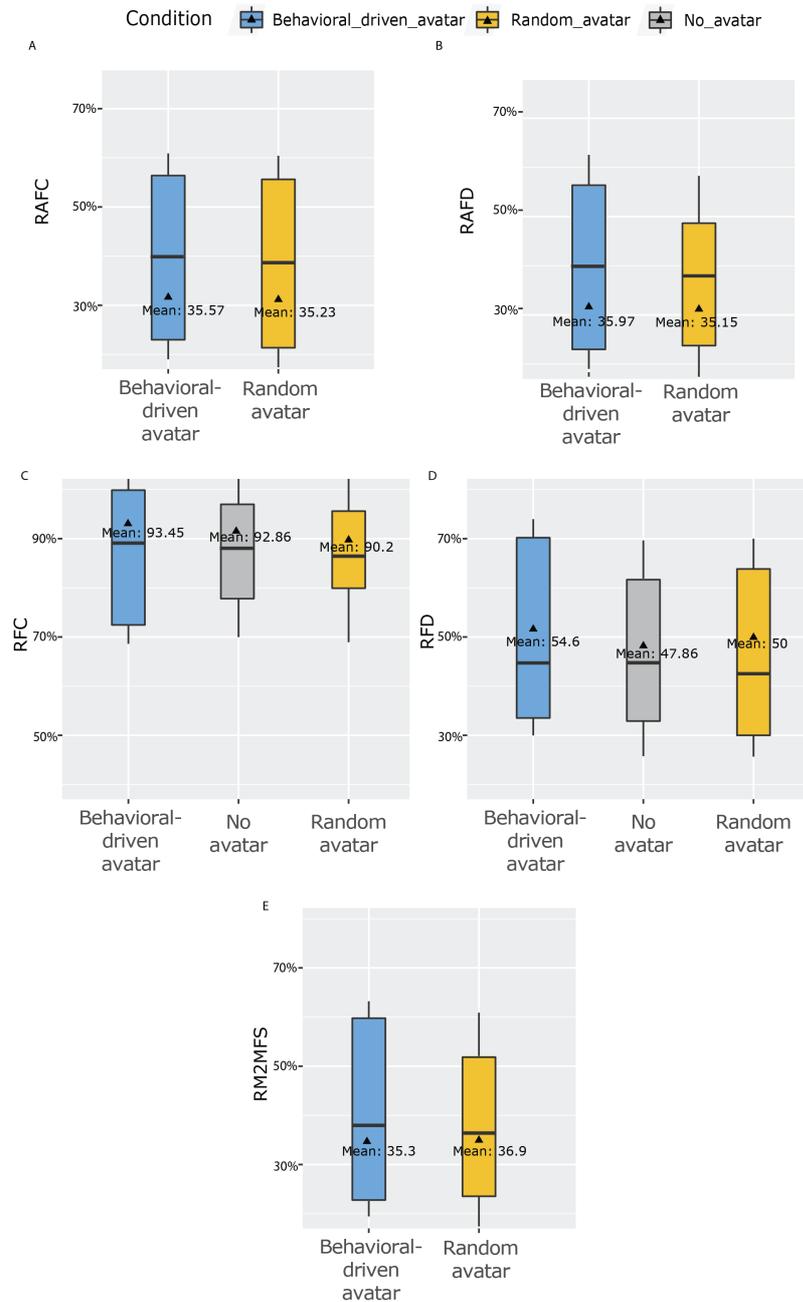

Fig. 12. Mean relative gaze behavioral data across conditions when having a formal discussion.

Taken together, the results indicated no significant differences between the behavior-driven avatar condition, the random avatar condition, and the no avatar condition in terms of gaze fixations on the AoI. This suggested that the avatar's presence and behavior did not significantly affect participants' fixation patterns. However, the analysis on RAS demonstrated that triads using our proposed system (SealMates) exhibited higher audio engagement from all participants than using a random avatar and no avatar in casual and formal discussions. Hence, **H3a and H3b were partially supported**).

## 6 DISCUSSION

The following section presents a succinct summary of our findings (also see Table 1 for a tabular summary) and an in-depth discussion of avenues for future work.

### 6.1 Mediating Online Group Interaction with a Collective Behavior-Driven Avatar

*6.1.1 Enhancing Self-Disclosure for Casual Conversation (RQ1).* Our results demonstrated that participants perceived their remote partners increased self-disclosure when a behavior-driven avatar was presented in video conferencing (subsection 5.1). The avatar's moving direction was designed to change based on the collective engagement level, including the speech and gaze information SealMates detected from each distributed member. It is possible that the movement of the SealMates gave distributed members some materials or excuses to invite others to speak, thus enabling them to break the ice more easily in a casual conversation. The effect of SealMates on a perceptual level is important for the follow-up social interaction. A positive perception of the interaction determines the initiation of future interaction. Past studies have established that animated avatars enable people to be expressive when exchanging messages in text-based communication [36]. Our results further demonstrate that a behavior-driven avatar in video conferencing can also make people feel that their remote parties are expressive.

Notably, SealMates did not make much difference in a formal discussion. One possible explanation is that the party-planning task with the assigned role did not require participants to disclose much about themselves. On the one hand, the choice of a task may limit our exploration of the effect of a behavior-driven avatar on formal discussions. On the other hand, it is also possible that a behavior-driven avatar is ineffective for influencing self-disclosure in a formal setting. Further investigation using other formal discussion tasks may be required.

*6.1.2 Enhancing Perceived Equal Participation (RQ2).* In the absence of an avatar, participants' perceptions of engagement were significantly polarized, with more participants reporting that other members were either highly engaged or not engaged at all rather than perceiving equal participation (subsection 5.2). This finding suggests that by using a visual representation of engagement, participants may tend to form equal judgments about the level of involvement of their fellow group members. One possible explanation is that the behavior-driven avatar provides additional social cues for people to infer the status from the remote sides. Lacking mutual gaze, body language, and proximal information makes it challenging for distributed members to infer the intentions of each other [32]. The behavior-driven avatar affords distributed members to update their impression of the group with an extra channel. By observing the avatar's moving directions, participants may have adjusted their perceptions of others' engagement accordingly, resulting in a more balanced distribution of judgments. The avatar's presence may have facilitated a shared understanding of engagement within the group, enabling participants to form a positive impression of the group.

Past studies have introduced chatbots to be actively involved in a group conversation verbally and mediate the conversation to facilitate equal contribution [20, 44]. Different from the previous approach, we proposed a rather subtle approach and demonstrated that merely moving an avatar across the virtual windows can also change distributed members' *perceptions* of equal participation.

*6.1.3 Increasing Individual Oral Participation (RQ3).* Although our results indicated that the presence of an avatar did not lead to significant differences in equal participation in terms of oral contribution or visual attention distribution, we found that having a behavior-driven avatar increased the total oral participation of each user, simplified in subsection 5.3. This finding is consistent with the literature demonstrating that technology, such as virtual agents or robots, can facilitate interpersonal conversation. For example, a recent study revealed that an embodied robot could facilitate deep conversation between two strangers [53]. Although we did not design the behavior-driven avatar to prompt interpersonal conversation, our results suggested that the avatar's mere movement can facilitate each member's oral engagement.

Table 1. Summary table of the results

| RQ | Dependent Variables | Casual conversation | Formal discussion |
| --- | --- | --- | --- |
| RQ1: How does the use of a behavior-driven avatar impact self-disclosure in casual conversation and formal discussion? | Self-disclosure | Behavior-driven avatar >no avatar (subsection 5.1, Figure 8A, H1a supported) | Behavior-driven avatar ~= no avatar (subsection 5.1, Figure 8B, H1b, not supported) |
| RQ2: How does the use of a behavior-driven avatar affect the perception of equal participation among members in casual conversation and formal discussion? | Perceived Level of Perceived Equal Participation | Perceived unequal participation was found in no avatar condition, but not in the behavior-driven avatar condition (subsection 5.2, Figure 9, H2a partially supported) | Perceived unequal participation was found in no avatar condition, but not in the behavior-driven avatar condition (subsection 5.2, Figure 10, H2b supported) |
| RQ3: How does the use of a behavior-driven avatar affect the level of equal participation for each group in casual conversation and formal discussion? | Total Individual Gazing and Speaking Time | There were not enough data points for analyzing gazing data in casual conversation. | No significant difference was found for gazing data across three conditions (Figure 12) |
| | | Behavior-driven avatar caused more audio participation from every member than no avatar condition. (subsection 5.3, Figure 11A, H3a supported) | Behavior-driven avatar caused more audio participation from every member than no avatar condition. (subsection 5.3, Figure 11B, H3b supported) |

### 6.2 Design Implications for Avatar-Mediated Communication

Based on current results, we provide two implications when designing video conferencing systems for avatar-mediated communication. First, consider *conversational contexts* when designing the avatar's action for mediating online social interaction. As our results revealed that people reacted differently to the avatar for casual conversation and formal discussion, designing avatar activity with contextual relevance to the video-conference objectives is essential. For example, an ice-breaking activity could feature avatar interactions that promote self-disclosure - such as signaling participants for introductions - and foster a positive environment with supportive and comforting gestures. Whereas a meeting with the objective of decision making, such as a vote, should feature avatar behavior that is more instrumental to the process: counting votes, timekeeping for speakers, or visualizing key points.

The second implication from our results is that future communication-mediating avatars should explore opportunities to augment group perception subtly. In contrast to other research that focuses on using a variety of chatbots to mediate group discussions proactively, there is much room for future investigation of biosignal-driven avatars that use a more nuanced approach to mediate remote conversations.

### 7 LIMITATIONS AND FUTURE WORK

Though SealMates was able to positively impact the online conversation, we also acknowledge some of the limitations present in this work. Firstly, the algorithm currently requires only two inputs; participants' gaze and speech pattern. Though these two features are a good indication of attention and participation respectively, there are other potential features that can be leveraged to understand their engagement. For example, previous works have explored the use of electroencephalography [25], galvanic skin response [38], and heart rate [15] to quantify attention. Our justification is that we do not want our participants to be encumbered by any additional sensing wearable given that we are proposing SealMates for a video conference platform. Yet, the use of eye gaze and speech presents its own limitations. Most notably, eye gaze requires minimal head movement to work effectively [26] and speech detection means that it can only function well in quiet environments [40]. In the future, we will expand the algorithm to not only be more robust to noise but to also include other forms of sensing.

For short-duration conversations such as our casual conversation which was six minutes, We could not obtain proper gaze results due to insufficient post-data-processing data to compare fixations and saccades, highlighted in subsection 5.3. Since our analysis was based mostly on the negotiation task setting, a limitation of our results was the impact of the different cultural backgrounds of participants, who may speak more or less to convince their peers of an objective. Some systematic limitations include time-windows of unusable data such as the first minute in Figure 5, and there was also the concern of users who became too aware of the system's behavior formula and attempted to adjust their behavior to influence the avatar. In future experiment designs, we can consider the option of different communication tasks, such as problem-solving, team-building, brainstorming, and other workplace-related scenarios[5].

We would like to remind the reader that SealMates only accounts for circumstances when all participants have their camera turned on, and any limitations when the camera is turned off fall outside the scope of this study. Future work should consider experimenting with SealMates while participants' cameras are off.

We acknowledge that all the survey responses provided by participants were based on self-reporting data, which may be subject to self-confirmation bias. There are numerous ways where future studies could considered to address this limitation, see [51]. However, it is important to note that self-reporting remains one of the most effective ways to capture attitudes and is commonly accepted in research as it provides a valuable perspective on participants' experiences [21]. Therefore, in the context of this research, we deemed it a suitable method. Additionally, we did not consider gender differences in the context of this research. Future research should take this factor into consideration as suggested by e.g. [45, 48]. Furthermore, we did not employ counterbalancing techniques when instructing participants to complete the casual conversation and formal discussion. Consequently, the potential influence of task order remains unknown. Future research should consider this aspect.

Lastly, SealMates is only proposed for conventional video conferencing platforms, and the algorithm was specifically designed to use speech and gaze data. We plan to further develop this collective behavior-driven avatar to be functional for other collaborative platforms like Meta Horizon Workrooms[6] that are purely on virtual reality (VR) or a telepresence robot. We will modify the collective behavior-driven avatar to accommodate various social interactions on these platforms, including other sensory feedback, body movement, etc. We only explored two main behaviors for the avatar: movement around the screen and change of animation and color. Yet, previous works have shown that other avatar representations like size [11], facial expression [29, 39], and shape [4] can also influence users.

## 8 CONCLUSION

We proposed SealMates, a behavior-driven avatar present on a video conference platform that infers the participants' participation and attention in the conversation. This was achieved using our developed algorithm that collected gaze and speech data and used it to move the avatar around the interlocutors' windows, as well as perform various animations to mediate the conversation. We conducted an experiment to determine its effect on the conversation by comparing three types of avatar interaction: behavior-driven avatar, random avatar, and no avatar. We found that our proposed behavior-driven allowed the participants to disclose themselves more and made the conversation more equally engaging for everyone. We also found higher fixation counts for users who participated and spoke more. We provided insights into these results and proposed potential future works to further expand SealMates, a collective behavior-driven avatar, to future collaborative platforms.

---

[5]https://www.betterup.com/blog/types-of-meetings
[6]https://www.oculus.com/workrooms/

## ACKNOWLEDGMENTS

This work is supported by JST Moonshot R&D Program (Grant number JPMJMS2013).